\documentclass[showpacs,amsmath,amssymb,aps,twocolumn,superscriptaddress,prx, footinbib]{revtex4-2}

\usepackage{tikz}
\usetikzlibrary{matrix,decorations.pathreplacing}
\usepackage{graphicx}   

\usepackage{mathtools}
\usepackage{algorithm}
\usepackage{amsthm}
\usepackage{amsmath}
\DeclareMathOperator{\Tr}{Tr}
\usepackage{algpseudocode}
\usepackage{grffile} 

\usepackage{ulem} 
\usepackage{comment}

\usepackage{float}
\usepackage{dcolumn}
\usepackage{bm}
\usepackage[colorlinks=true]{hyperref}

\hypersetup{
	colorlinks=true,
	linkcolor=blue,
	urlcolor=cyan,
}

\usepackage{color}
\definecolor{ForestGreen}{RGB}{34, 139, 34}

\newcommand{\affA}{School of Physics, Peking University, 100871 Beijing, China.}
\newcommand{\affB}{Max Planck Institute for the Physics of Complex Systems, N\"othnitzer Str.~38, 01187 Dresden, Germany.}
 \newcommand{\affC}{Department of Physics, Princeton University, Princeton, New Jersey 08544, USA}

\begin{document}
		
\title{Non-Hermitian delocalization in 1D via emergent compactness}

\author{Liang-Hong Mo}
 \affiliation{\affA}
 \affiliation{\affB}
  \affiliation{\affC}
   
\author{Zhenyu Xiao}
 \affiliation{\affA}
  
\author{Roderich Moessner}
 \affiliation{\affB}

 	\author{Hongzheng Zhao}
	\email{hzhao@pku.edu.cn}
 \affiliation{\affA}
    \affiliation{\affB}
	
	\date{\today}

\begin{abstract}
Potential disorder in 1D leads to Anderson localization of the entire spectrum. Upon sacrificing Hermiticity by adding non-reciprocal hopping, the non-Hermitian {skin effect} competes with localization. We find another route for delocalization, which involves {\it imaginary} potential disorder. While an entirely random potential generally still leads to localization,  imposing minimal spatial structure to the disorder can protect delocalization: it endows the concomitant transfer matrix with an $\mathrm{SU}(2)$ structure, whose compactness in turn translates into an infinite localization length. The fraction of delocalized states can be tuned by the choice of boundary conditions.

\end{abstract}
\maketitle
\let\oldaddcontentsline\addcontentsline
\renewcommand{\addcontentsline}[3]{}

\textit{Introduction.---} 
{Anderson localization~\cite{Anderson58,2008RvMP...80.1355E}}
is an everlasting subject in condensed matter physics. The presence of quenched disorder significantly suppresses transport, especially in low dimensions. This stimulates decades of studies including, e.g., delocalization phase transitions with or without many-body interaction~\cite{nandkishore2015many,smith2017disorder,abanin2019colloquium}, and topological phases of matter in disordered systems~\cite{li2009topological,titum2016anomalous,stutzer2018photonic}.
 
Recently, localization in open systems has attracted considerable attention~\cite{hatano1996localization, Feinberg97,Feinberg99,regensburger2013observation,Xu16,ding2016emergence,martinez2018non, Gong18,hamazaki2019non,Tzortzakakis20,Wang20,Huang20,borgnia2020non,Huang20SR,Kawabata20,Luo21,Luo21TM,luo2022,zhu2022hybrid,li2022gain,zhang2023emergent,manna2023inner,zhu2023higher,thompson2024,li2024universal,xing2024universal, halder2024competing}. The interplay between non-Hermicity and localization leads to a plethora of novel phenomena, unique in open quantum systems~\cite{yao2018edge,yokomizo2019non,okuma2020topological,ashida2020non}. As shown in Hatano and Nelson’s pioneering work~\cite{hatano1996localization}, in one-dimension (1D) asymmetric hopping can compete with quenched disorder, leading to delocalization, at variance with the scaling theory of localization for potential disorder~\cite{2008RvMP...80.1355E}. 
Such delocalization originates from the non-trivial topology~\cite{Gong18,okuma2023non}, as elucidated by the effective field theory~\cite{Kawabata20,chen2024}.
Also, by introducing a deterministic quasi-periodic spatial structure delocalization may also appear~\cite{longhi2019topological,liu2021localization,zhai2021cascade,chen2022quantum}, similar to their Hermitian counterpart~\cite{aubry1980analyticity,liu1987electronic,axel1989spectrum,biddle2009localization,wang2020one}.

Here, we address the intriguing question of whether 1D delocalized eigenstates can be realized by purely on-site energy gain and loss, without sacrificing reciprocity and spatial randomness. 
In general, a purely random complex potential localizes the entire spectrum, although prior work demonstrates the possibility of exotic transport dynamics in wave propagation
~\cite{basiri2014light,yusipov2017localization,tzortzakakis2021transport,weidemann2021coexistence,li2024universal}. 

\begin{figure*}[t]
\centering
\includegraphics[width=\linewidth]{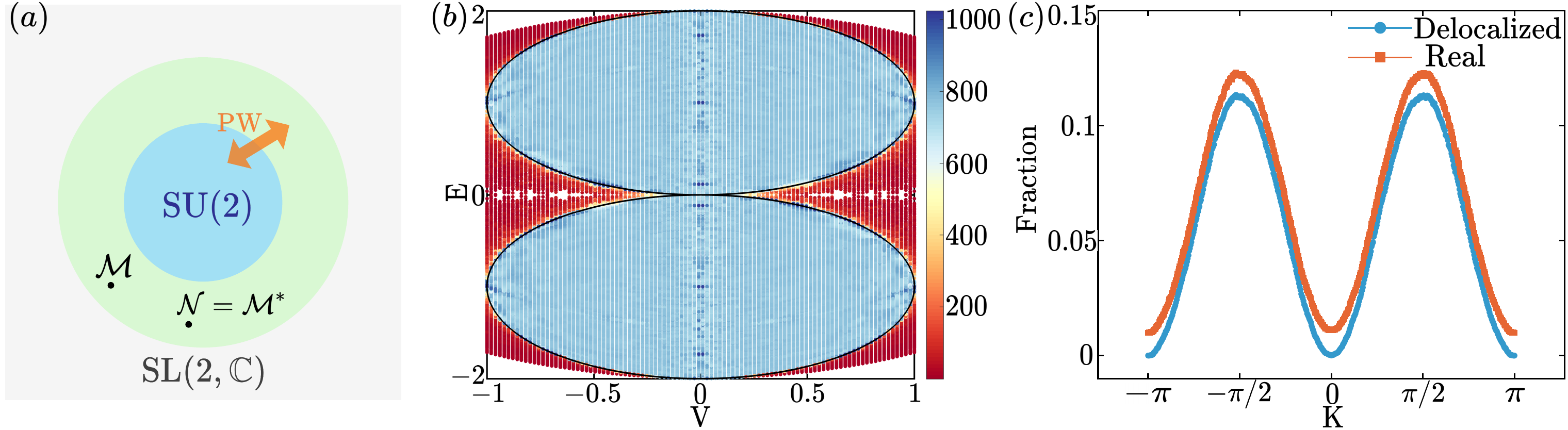} 
\caption{(a) Schematic for the emergent compact structure. The overall gray region denotes the non-compact $\mathrm{SL}(2,\mathbb{C})$ group, which contains the compact $\mathrm{SU}(2)$ structure (blue). The green region denotes a subset of $\mathrm{SL}(2,\mathbb{C})$, inside which a matrix element as well as its conjugation can be simultaneously mapped to $\mathrm{SU}(2)$ matrices, via a similarity transformation PW. The dipolar transfer matrices, $\mathcal{M}$ and $\mathcal{N}$, belong to this subset and hence a random product of the two leads to zero Lyapunov exponent, i.e., an infinite localization length. (b) Participation ratio (PR) calculated for each eigenstate at different real energies $E$ and disorder strength $V$. The emergent compactness leads to delocalized eigenstates (blue), which are separated from the localized states (red) by an exact mobility edge (black). This black line is obtained by analyzing the possible existence of a similarity transformation. In numerical simulation, we scan over all possible boundary conditions with a twisted angle $K$ and sample different realizations of the disorder potential. (c) 
By tuning $K$, we can control the fraction of real energies (orange) in the entire spectrum and the delocalized eigenstates therein (blue). {For periodic and anti-periodic boundary conditions $K{=}0, \pi$, the fraction of eigenstates is around $10^{-4}$, a small but nonzero value.} Yet, for other $K$ values, most real energies are delocalized. For numerical simulation, we use $V=0.3$, $2L=1024$. We average over $100$ random realizations to obtain both (b) and (c). 
 }
\label{Fig:PWmapping+Percentagerealspectra+IPRdiagram}
\end{figure*}

We provide an affirmative answer by constructing a concrete model with delocalized eigenstates and analytically tractable mobility edges. The key ingredient involves imposing a minimal structure on a binary potential, which we dub ``dipolar disorder". The corresponding transfer matrix generally belongs to the non-compact $\mathrm{SL}(2,\mathbb{C})$ group~\cite{hall2013lie,woit2017quantum}; hence, a random product of such matrices leads to a positive Lyapunov exponent, i.e., a finite localization length~\cite{evers2008anderson}. 
However, the possible existence of an emergent compact $\mathrm{SU}(2)$ structure, which we identify, in the transfer matrix for the real-valued spectrum results in a zero Lyapunov exponent. 
It, in turn, leads to an infinite localization length, protecting delocalization against disorder. This $\mathrm{SU}(2)$ structure can be revealed by a simple analytical procedure, cf.~Fig.~\ref{Fig:PWmapping+Percentagerealspectra+IPRdiagram}(a), allowing the exact determination of a mobility edge, for which the usual numerical approach faces difficulties, especially in the thermodynamic limit.

We numerically verify the theoretical prediction by diagonalizing the corresponding lattice model. 
A natural diagnostic of delocalization is the participation ratio of eigenstates, by which we obtain a localization-delocalization phase diagram at different energies, Fig.~\ref{Fig:PWmapping+Percentagerealspectra+IPRdiagram}(b). The mobility edge precisely matches the analytical prediction. Crucially, the fraction of delocalized eigenstates depends on the boundary conditions, which are tunable by a twisted phase factor, cf. Fig.~\ref{Fig:PWmapping+Percentagerealspectra+IPRdiagram}(c). For complex energy eigenvalues that go beyond this analytical framework, eigenstates are generally localized. However, for a given finite system size, we find strong numerical evidence showing that delocalization can persist as long as the imaginary component of the energy eigenvalue is sufficiently small.

The remainder of this account is organized as follows. First, we introduce the model with disordered imaginary potential and the dipolar structure. 
Then we elaborate on the emergent $\mathrm{SU}(2)$ structure in the transfer matrix. 
By numerically calculating the participation ratio of each eigenstate, we demonstrate the existence of delocalization and the localization transition. Finally, we summarize and discuss open directions where the emergent compactness in the transfer matrix may play a crucial role.  

\textit{The model.---}
The Schrödinger equation for 1D tight-binding Hamiltonian $H$ with uniform and reciprocal nearest-neighbor hopping and on-site potential $V_j$ reads
\begin{align} \label{eq:Schrodinger}
    E\Psi_j=\Psi_{j+1}+\Psi_{j-1}+V_j\Psi_j, \quad j\in \mathbb{Z},
\end{align}
where \(\Psi_j\equiv \langle j|\Psi\rangle\) is the real-space wavefunction on site \(j\) and $E$ defines the energy. 
It can be rewritten as
\begin{align}
    \begin{pmatrix}
\Psi_{j+1} \\
\Psi_{j}
\end{pmatrix}=T_j\begin{pmatrix}
\Psi_{j} \\
\Psi_{j-1}
\end{pmatrix}, \ T_j=\begin{pmatrix}
E-V_j  &-1 \\
 1 &0
\end{pmatrix},
\end{align}
where $T_j$ is the transfer matrix. For the entire chain with length $N$, the transfer matrix reads 
$T^{(N)}_{\text{tot}}{=}\prod_{j=1}^{N}T_j$.  The corresponding Lyapunov exponent $\lambda_L$ is defined as $ \lambda_L{\equiv} \lim\limits_{N\rightarrow \infty} 
{\log||T^{(N)}_{\text{tot}}||}/{N} $, where $||...||$ denotes the $2$-norm~\footnote{The corresponding Lyapunov exponents $\lambda_i$ ($i=1,2$; $\lambda_1 \geq \lambda_2$) are defined by the eigenvalues of $\lim_{N \rightarrow \infty } \log [(T^{(N)}_{\text{tot}})^{\dagger} T^{(N)}_{\text{tot}}]/2N $.
The localization length is defined as $\xi(E) \equiv (\min_i |\lambda_i|)^{-1}$ for an eigenstate with energy $E$.
For the $2\times 2$ transfer matrix here, we always have $\lambda_2 = -\lambda_1$, and hence $\xi = \lambda_1^{-1}$. It is easy to prove that $\lambda_1$ is equivalent to the Lyapunov exponent defined through $2$-norm.
}. For a given eigen-energy $E$,  the localization length $\xi (E)$ can be obtained as the inverse of the Lyapunov exponent $\lambda_L$ of $T^{(N)}_{\text{tot}}$~\cite{carmona1982exponential,comtet2013lyapunov,slevin2014}.
In the conventional Hermitian Anderson model where $V_j$ is a real random variable, 
it has been shown that $\lambda_L$ is positive for any energy $E$, therefore all eigenstates of the 1D Anderson model are localized.

Now we consider a non-Hermitian model with imaginary and binary potential, i.e., $V_j{=}\pm iV$, $V{\in} \mathbb{R}$. We focus on states with real energies and obtain two transfer matrices
\begin{align} T_{\pm}=\begin{pmatrix}
E\pm iV &-1 \\
 1 &0
\end{pmatrix},
\label{eq:Tmatrix}
\end{align}
which satisfy $T_{+}=T_{-}^*$ and belong to the group $\mathrm{SL}(2,\mathbb{C})$. Its associated manifold is non-compact~\cite{hall2013lie,woit2017quantum}. According to Furstenberg's theorem~\cite{furstenberg1963noncommuting}, 
in a non-compact manifold, a random product of matrices generally leads to a positive Lyapunov exponent. 
Therefore, just as the conventional Hermitian Anderson model, 
the onsite imaginary potential can localize all eigenstates, in accordance with previous predictions in Refs.~\cite{basiri2014light,weidemann2021coexistence,li2024universal}. 

\textit{Emergent compactness.---}
We impose a dipolar structure into the spatial disorder, such that the minimal building block of the disorder becomes either $\{iV, -iV\}$, or $\{-iV, iV\}$. 
{Each block has a balanced energy gain ($iV$) and loss ($-iV$), resulting in a parity-time symmetry within the block, also referred to as ``local parity-time symmetry"~\cite{bendix2010optical}.
By contrast, global parity-time symmetry is absent due to the spatial randomness of the blocks.}
The corresponding transfer matrices for two consecutive sites become $\mathcal{M}{=}T_+T_-,\ \mathcal{N}{=}T_-T_+$, which satisfy $\mathcal{M}=\mathcal{N}^*$ and exhibit a real trace.

A peculiar feature appears in this dipolar construction:
as schematically illustrated in Fig.~\ref{Fig:PWmapping+Percentagerealspectra+IPRdiagram}(a), an emergent compactness arises because both $\mathcal{M}$ and $\mathcal{N}$ can simultaneously be transformed into $\mathrm{SU}(2)$  matrices via a similarity transformation. As the corresponding manifold for the $\mathrm{SU}(2)$ group is compact, the Lyapunov exponent for a series of random multiplication of $\mathcal{M}$ and $\mathcal{N}$ vanishes. 

A subset of real eigenenergies allows for such a similarity transformation and the crucial ingredient is simply: 
\begin{equation}
\label{eq:tracecondition}\Tr{(\mathcal{M}\mathcal{N})}\le 2.
\end{equation}This, together with the condition $\mathcal{M}=\mathcal{N}^*$, guarantees that the eigenvalues of both \( \mathcal{M} \) and \( \mathcal{N} \) satisfy the spectral requirements of the $\mathrm{SU}(2)$ group. Numerically, one can efficiently scan over the entire parameter space and show that Eq.~\ref{eq:tracecondition} leads to the blue region in Fig.~\ref{Fig:PWmapping+Percentagerealspectra+IPRdiagram}(b), including the black line~\footnote{The analytical expression for the mobility edge is $
- 4E^2 + E^4 + 2E^2 V^2 + V^4= 0
$}.

To construct the desired similarity transformation, we can first diagonalize the matrix \( \mathcal{M} \) via a {normalized} matrix $P$, which has the following general form $P=\begin{pmatrix} 
a & b \\ 
c & d 
\end{pmatrix}. $ One can show that Eq.~\ref{eq:tracecondition} ensures that the entire blue region satisfies $(b^*d - bd^*)(-a^*c + ac^*)\le 0$. When the equal sign is taken, the matrix $P$ is sufficient to simultaneously transform both $\mathcal{M}$ and $\mathcal{N}$ into two $\mathrm{SU}(2)$ matrices. Otherwise, one can construct a product of matrix $P$ and $W$, where \begin{align}
    W=
 \begin{pmatrix}
1 & 0 \\
0 & \sqrt{\frac{-a^*c + ac^*}{b^*d - bd^*}}
\end{pmatrix},
\end{align}
to achieve such a similar transformation. A detailed proof and a more general discussion is presented in the Appendix.

Consequently, inside the blue region in Fig.~\ref{Fig:PWmapping+Percentagerealspectra+IPRdiagram}(b), the dipolar transfer matrices have an emergent compact $\mathrm{SU}(2)$ structure, resulting in a vanishing Lyapunov exponent and a diverging localization length. In contrast, for parameters in the red region in Fig.~\ref{Fig:PWmapping+Percentagerealspectra+IPRdiagram}(b), both $\mathcal{M}$ and $\mathcal{N}$ have an eigenvalue larger than one, thus they do not belong to any compact subgroup of $\mathrm{SL}(2,\mathbb{C})$. Therefore, localization is expected.

\textit{Numerical results.---}
It is worth highlighting that, here the energy eigenvalue $E$ is not a continuously tunable parameter. 
Rather,  it depends on the specific spatial disorder realization as well as the boundary condition of the lattice. In the following, we show that a nonzero fraction of real eigenvalues, as well as delocalized eigenstates therein, can exist, and can be tunable by the twisted boundary condition.

For a system containing $2L$ sites ($L$ dipoles), the spectrum can be obtained by solving the following bulk equation and boundary condition
\begin{align}
\label{eq:bulk-boundary-equation}
&\begin{pmatrix}
\Psi_{2L+1} \\
\Psi_{2L}
\end{pmatrix}=T^{(2L)}_{\text{tot}}\begin{pmatrix}
\Psi_{1} \\
\Psi_{0}
\end{pmatrix}, \ \begin{pmatrix}
\Psi_{2L+1} \\
\Psi_{2L}
\end{pmatrix}=e^{iK}\begin{pmatrix}
\Psi_{1} \\
\Psi_{0}
\end{pmatrix},
\end{align}
leading to 
\begin{align}
\label{eq:bulk-boundary-equation2}
T^{(2L)}_{\text{tot}}\begin{pmatrix}
\Psi_{1} \\
\Psi_{0}
\end{pmatrix}=e^{iK}\begin{pmatrix}
\Psi_{1} \\
\Psi_{0}
\end{pmatrix},
\end{align}
where we choose twisted boundary condition with a real number $K$~\cite{kohmoto1986localization}. $K{=}0$ corresponds to a periodic boundary condition. 

If $\mathcal{M}$ and $\mathcal{N}$ can be simultaneously transformed into $\mathrm{SU}(2)$ matrices, the eigenvalues of \(T_{\text{tot}}\) take the form of a \(U(1)\) phase factor. Therefore, for a specific set of $V$ and $E$, there exists a suitable boundary condition $K$ such that the energy $E$ exists in the spectrum.
\begin{figure}[t]
\centering
\includegraphics[width=0.9\linewidth]{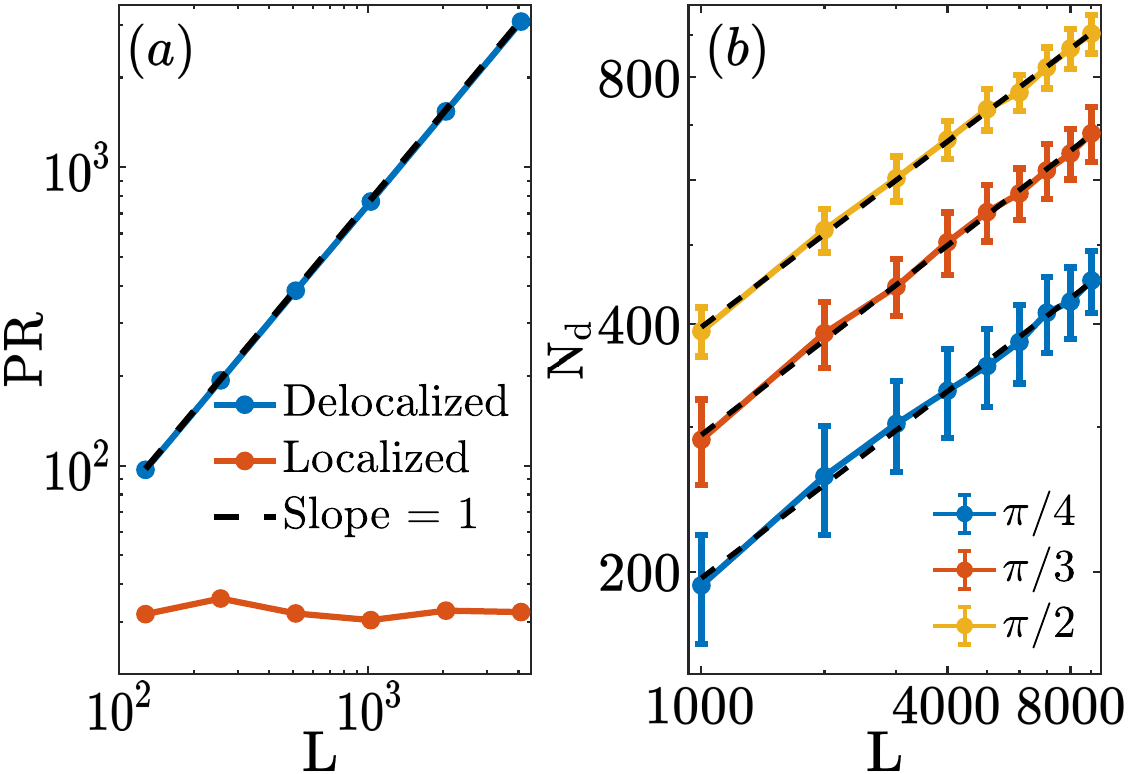} 
\caption{(a) Scaling of the participation ratio (PR) for different system sizes. For eigenstates in the blue region of Fig.~\ref{Fig:PWmapping+Percentagerealspectra+IPRdiagram} (b), PR scales linearly with system size (black dashed line of slope 1 is a guide to the eye), confirming its delocalized nature. In contrast, for the eigenstates in the red region of Fig.~\ref{Fig:PWmapping+Percentagerealspectra+IPRdiagram}(b), PR is independent of system size, corresponding to localized states. We use the parameters $V=0.5$ and $K=\pi/2$ for numerical simulation. (b) Finite-size scaling of the number of delocalized eigenstates with real energies, $N_{\mathrm{d}}$, averaged over $100$ random disorder realizations. Error bars represent statistical fluctuations, and different colors correspond to distinct \(K\) values. $N_{\mathrm{d}}$ appears to grow algebraically with system size, with the black dashed line representing $L^{0.4}$ as a guide to the eye. We set \( V = 0.1 \) for numerical simulation. 
}
\label{Fig:IPRscaling}
\end{figure}

\begin{figure}[t]
\centering
\includegraphics[width=\linewidth]{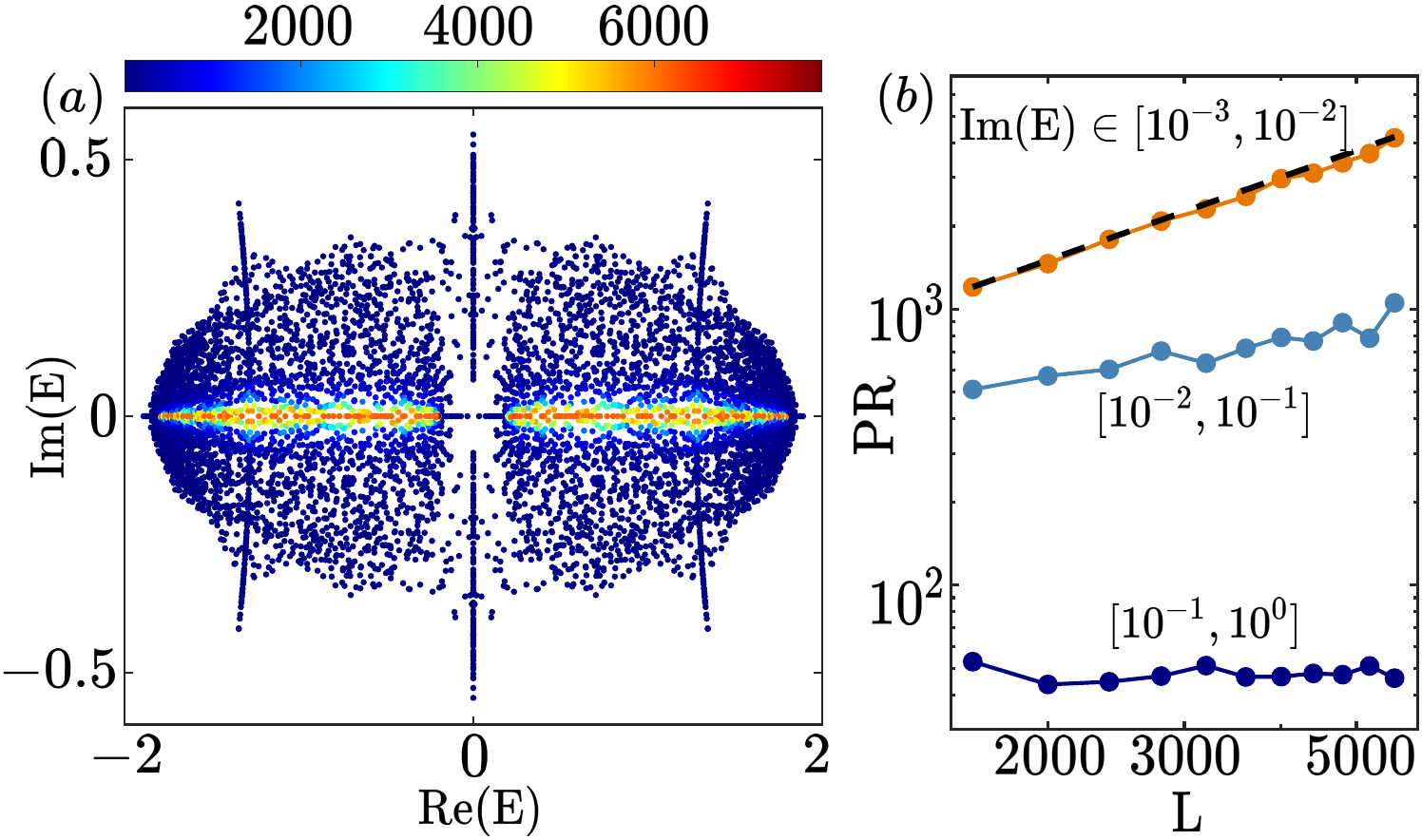} 
\caption{(a) PR for the entire complex spectrum. Our simulations use \( V = 0.6, K=\pi/2 \) and system size \( L = 8192 \). Most eigenstates with complex energies are localized, while a small fraction of energies near the real axis (bright region) exhibits large PR values, suggesting delocalization may persist. (b) System size scaling of PR, averaged over eigenstates within different imaginary energy windows. The
real part of the energy is fixed within the range \([0.8,1.2]\). For an eigenvalue sufficiently close to the real axis, PR exhibits linear scaling in $L$ (the black dashed line of slope 1 is a guide to the eye). When deviating further away from the real axis, a power law scaling appears with a scaling exponent smaller than 1. A log-log scale is used here.
}
\label{Fig:PRofspectrum}
\end{figure}

To verify this, we numerically perform exact diagonalization of the Hamiltonian
\begin{eqnarray}
    \begin{aligned}
    \label{eq:Hamiltonian}
\mathcal{H}{=}&\sum_{j=0}^{2L-2} (\hat c_j^{\dagger} \hat c_{j+1} + {h.c.}) {+} \sum_{j=0}^{L-1} \left( V_j \hat c_{2j}^{\dagger} \hat c_{2j} {-} V_j \hat c_{2j+1}^{\dagger} \hat c_{2j+1} \right)\  \\
    &+ \left[\exp(iK)\hat c_0^{\dagger} \hat c_{2L-1} + {h.c.}\right],
\end{aligned}
\end{eqnarray}
and obtain the entire single-particle energy eigenvalues for different $K$. 
As expected, since the Hamiltonian is non-Hermitian, energy eigenvalues {are mostly complex}, see Fig.~\ref{Fig:PRofspectrum} (a). 
However, crucially, real eigenvalues also exist and their fraction depends on the boundary condition: As depicted in Fig.~\ref{Fig:PWmapping+Percentagerealspectra+IPRdiagram} (c), for periodic and anti-periodic boundary conditions (\(K {=} 0\) and \(\pi\)) {this fraction (orange) is around $10^{-2}$, and the fraction for delocalized eigenstates (blue) is around $10^{-4}$, see details in the Supplementary Material (SM), Sec.~\ref{SM:percentage}.} 
At the same time, for other $K$ values, a notable fraction of real eigenvalues appears, and most of the corresponding eigenstates are indeed delocalized. Note, these results can be efficiently obtained by counting the number of states located inside the black line in Fig.~\ref{Fig:PWmapping+Percentagerealspectra+IPRdiagram} (b). 

We further perform a systematic finite-size scaling analysis of the delocalized eigenstates. As shown in Fig.~\ref{Fig:IPRscaling}(b), for three different $K$ values, the number of delocalized eigenstates, $N_{\mathrm{d}}$, increases with system size. This dependence is consistent with a power law over the range of accessible system sizes, $L^{\alpha}$, where the scaling exponent $\alpha$ can be fitted numerically. Such a power law dependence also appears for other disorder strengths $V$.  Therefore, $N_{\mathrm{d}}$ diverges in the thermodynamic limit. The scaling exponent $\alpha$ can vary for different $V$, typically ranging from 0.2 to 0.4, see further numerical results in Sec.~\ref{SM:finitesize}. However, for the boundary conditions $K=0$ and $\pi$, $N_{\mathrm{d}}$ does not grow with system size, resulting in the small fraction as shown in Fig.~\ref{Fig:PWmapping+Percentagerealspectra+IPRdiagram} (c) when $L$ is large.

To characterize the localization properties of the system, we calculate the participation ratio (PR)~\cite{longhi2019topological} for each normalized right eigenvector $|\Psi\rangle$,
\begin{align}
    \text{PR} = \left(\sum_{j=0}^{2L-1} |\langle j|\Psi\rangle|^4\right)^{-1}.
\end{align}
For localized states, PR is of order $\mathcal{O}(1)$ and remains independent of the system size; for delocalized states, it becomes proportional to the system size, $\mathcal{O}(L)$. Fig.~\ref{Fig:PWmapping+Percentagerealspectra+IPRdiagram}(b) presents the PR for all eigenstates with real eigenvalues, scanning over a wide range of the disorder strength $V$ and different boundary conditions $K$. In the absence of disorder ($V{=}0$), all states are delocalized in space, exhibiting large values of $\text{PR}$ (blue). For non-zero $V$, localized eigenstates appear with a notably smaller $\text{PR}$ (red). Importantly, a mobility edge, separating the localized and delocalized regimes, is clearly visible and it precisely matches the analytical prediction (black line).

In Fig.~\ref{Fig:IPRscaling}(a), we depict the PR averaged over either the localized states (orange) or the delocalized states (blue) for different system sizes, where the linear system size dependence in PR confirms delocalization.

\textit{Complex spectrum.---}
The delocalized eigenstates and their localization transition have been well-established for real energies. For complex energies, the existence of the emergent $\mathrm{SU}(2)$ structure is not guaranteed, nor is the existence of a delocalized eigenstate. In fact, most of the complex spectrum becomes localized, as shown in Fig.~\ref{Fig:PRofspectrum} (a), where PR for the entire spectrum is plotted. 

For eigenvalues close to the real axis, matrix elements of $T_{\pm}$ (Eq.~\ref{eq:Tmatrix}) are still dominated by the real part of the energy. Therefore, despite the energy eigenvalues not being purely real, the resulting localization length can still be sufficiently large, even comparable to system sizes that are numerically accessible. Indeed, as shown in Fig.~\ref{Fig:PRofspectrum} (a), a bright region with large PR values appears, indicating the possible existence of delocalized eigenstates. 

A systematic analysis of the system size dependence is presented in Fig.~\ref{Fig:PRofspectrum} (b). The real part of the energy is fixed within the range $\text{Re}(E){\in}[0.8,1.2]$, such that all eigenstates are delocalized if their energies are purely real. We average the PR over all eigenstates within different imaginary energy windows.
For eigenstates with a small imaginary part, $\text{Im}(E){\in} [10^{-3},10^{-2}]$ (orange), {the averaged PR grows linearly with $L$, indicating that delocalization persists even for a complex eigenvalue. This can be a finite-size effect, but the localization length in practice is comparable to or larger than the current system size.} 
Consider $\text{Im}(E){\in} [10^{-2},10^{-1}]$ (light blue), where a crossover behavior appears, showing
a power law scaling with a scaling exponent less than $1$. In contrast, for $\text{Im}(E){\in} [0.1,1]$ (dark blue), PR becomes independent of system size, implying localization.

\textit{Discussion.---}
We have presented a simple 1D non-Hermitian system with a purely imaginary disorder potential that leads to delocalized eigenstates. 
The crucial ingredient is an emergent compact $\mathrm{SU}(2)$ structure in the concomitant transfer matrix. It leads to a zero Lyapunov exponent and hence an infinite localization length for real eigenvalues, as further supported by numerical simulation. For complex energies, delocalization persists as long as the imaginary component of the eigenvalue is sufficiently small. It would be worthwhile to perform further numerical investigations to clarify whether this is a finite-size phenomenon.

We also note that the spectrum exhibits an interesting $D_2$ symmetry, namely the mirror symmetry w.r.t. both real and imaginary axes. For the imaginary axis, the symmetry originates from a chiral symmetry,
$
P{H}^{\dagger}P^{-1}=-{H},
$
where ${H}$ is the single-particle Hamiltonian matrix of Eq.~\ref{eq:Hamiltonian} and $P$ denotes the chiral symmetry operator. This leads to the coexistence of the energies $E$ and $-E^*$.
Additionally, one can show that energy eigenvalues also appear in pairs, $E$ and $-E$. This is more subtle and can be revealed by showing that the trace of the total transfer matrix only contains even powers of the energy $E$ for twist boundary conditions. Details of the proof are illustrated in Sec.~\ref{Sec:symmetry}. The connection between this symmetry and delocalization is
an intriguing subject for future study. 

For disordered higher dimensional systems, a generic transfer matrix of large dimension normally appears, and the possible existence of a compact structure remains an open and interesting question. 

{While focusing on the eigenstate properties of the system, we emphasize that this emergent compact structure can also lead to interesting dynamical phenomena that may not exist in non-Hermitian systems with an entirely localized spectrum}~\cite{zheng2010pt,regensburger2012parity,li2024universal,xing2024universal}.

There are various experimental platforms capable of realizing tunable control over the on-site gain and loss, {such as photonic waveguides~\cite{guo2009observation,cong2021temporal,lin2022topological,ma2024anisotropic} and mechanical systems~\cite{susstrunk2016classification}}.
Therefore, the proposed dipolar potential structure, as well as the predicted delocalization should be readily accessible in these physical settings.

Finally, we emphasize that, due to the broad applicability of the transfer matrix method, the emergent compactness could have significant implications at a more general level, particularly in many-body systems. For instance, a similar transfer matrix method has been developed in the study of dual unitary and Gaussian quantum circuits~\cite{claeys2021ergodic,granet2023volume}, as well as within the framework of time-dependent driven systems exhibiting conformal symmetry~\cite{2020PhRvX..10c1036F}.
We anticipate that this emergent compactness will lead to more surprises and unexpected dynamical non-equilibrium many-body phenomena.

\textit{Acknowledgments.---}
We thank Kun Ding, Yumin Hu, Sen Mu, and Bo-Ting Chen for stimulating discussions. This work is supported by the National Natural Science Foundation of China (Grant No. 12474214), by “The Fundamental Research Funds for the Central Universities, Peking University”, by ``High-performance Computing Platform of Peking University" and by the Deutsche Forschungsgemeinschaft under the cluster of excellence ct.qmat (EXC 2147, project-id 390858490).

\bibliography{Reference}

\part*{\scalebox{0.6}{End Matter}}
\addcontentsline{toc}{part}{End Matter}

\textit{Appendix.---}

In the following, we prove that the condition $\mathrm{tr}(\mathcal{M}\mathcal{N})\in [-2,2]$ leads to an emergent compact $\mathrm{SU}(2)$ structure in the transfer matrix of our system.  

We first consider two arbitrary $\mathrm{SL}(2,\mathbb{C})$ matrices $\mathcal{M},\mathcal{N}$ and analyze the minimal conditions that allow for the existence of a similarity transformation to simultaneously transform them into two $\mathrm{SU}(2)$ matrices. The general form of a $\mathrm{SU}(2)$ matrix reads $\begin{pmatrix}
  \alpha &\beta \\
-\beta^*  &\alpha^*
\end{pmatrix}$. Then we apply this framework to our specific model being considered in the localization problem. These conditions read:
\begin{enumerate}
    \item $\mathcal{M},\mathcal{N}$ are not triangular matrices and they satisfy $ \mathcal{M}=\mathcal{N}^*$.
    \item $\mathrm{tr}(\mathcal{M})\in[-2,2]$.
    \item $   \mathrm{tr(} \mathcal{M}\mathcal{N})\le 2$.
    
\end{enumerate}

\begin{proof}

Condition 1 and $2$ imply that the eigenvalues of $\mathcal{M}$ and $\mathcal{N}$ are $u+iv,u-iv$ with $u^2+v^2=1.$ Thus they can be expressed as
\begin{align}
\label{eq:diagonalmatrix}
  &  \mathcal{M}=P\Lambda P^{-1},P=\begin{pmatrix}
a  & b\\
c  &d
\end{pmatrix},\Lambda=\begin{pmatrix}
u+iv  & 0\\
0  &u-iv
\end{pmatrix},
\\ &\mathcal{N}=P^*\Lambda^* (P^*)^{-1},
\end{align}
where each column of  $P$ is an eigenvector of $ \mathcal{M}$. For simplicity, we consider $P$ to be normalized and $\det(P)=1$.

By construction, $P$ diagonalizes $\mathcal{M}$ into $\Lambda$, which is a $\mathrm{SU}(2)$ matrix. Similarly, $P$ transforms $\mathcal{N}$ to
\begin{align}
    P^{-1}\mathcal{N}P=P^{-1}P^*\Lambda^* (P^*)^{-1}P,
\end{align}
which, in general, is not a $\mathrm{SU}(2)$ matrix,
and we have the explicit expression
\begin{align}
\label{eq:PP}
P^{-1}P^*=\begin{pmatrix}
-bc^* + a^*d & b^*d - bd^* \\
-a^*c + ac^* & -b^*c + ad^*
\end{pmatrix}.
\end{align}
Note, its diagonal elements are conjugates of each other, an important feature of a generic $\mathrm{SU}(2)$ matrix. However, its off-diagonal elements can break the $\mathrm{SU}(2)$ structure.
There are three possibilities:

Case 1: All off-diagonal elements are zero. \( P^{-1}P^* \) reduces to the identity matrix, which is naturally a $\mathrm{SU}(2)$ matrix; 

Case 2: Only one of the off-diagonal elements is non-zero. If so, $\mathcal{M}$ becomes a triangular matrix and hence Condition 1 is not satisfied.  

Case 3: Both off-diagonal elements are non-zero. In this case, we need to introduce the one extra matrix 
\begin{align}
    W=
 \begin{pmatrix}
1 & 0 \\
0 & \sqrt{\frac{-a^*c + ac^*}{b^*d - bd^*}}
\end{pmatrix}.
\end{align}

By applying a similarity transformation $W$ to Eq.~\ref{eq:PP}, we have
\begin{align}
    U&\equiv  W^{-1}P^{-1}P^*W\\\nonumber
&=\begin{pmatrix}
-bc^* + a^*d & \left(b^*d - bd^*\right) \sqrt{\frac{a^*c - ac^*}{-b^*d + bd^*}} \\
\left(b^*d - bd^*\right) \sqrt{\frac{a^*c - ac^*}{-b^*d + bd^*}} & -b^*c + ad^*
\end{pmatrix}.
\end{align}
By comparing $U$ with the general form of a $\mathrm{SU}(2)$ matrix, one finds that if and only if  $(b^*d - bd^*)(-a^*c + ac^*)< 0$, $U$ becomes an $\mathrm{SU}(2)$ matrix. 

Therefore, by using the matrix $PW$ one can transform $\mathcal{N}$ as
\begin{align}
(PW)^{-1}    \mathcal{N}PW&=
U\Lambda^*U^{-1},
\end{align}
which is now a product of three $\mathrm{SU}(2)$ matrices.

In summary, based on the first two requirements as well as $(b^*d - bd^*)(-a^*c + ac^*)\le 0$,  $\mathcal{M}$ and $\mathcal{N}$  can be simultaneously similarly transformed into $\mathrm{SU}(2)$ matrices.

In the following, we show that the condition $(b^*d - bd^*)(-a^*c + ac^*)\le 0$ is equivalent to a simpler expression 
\begin{align}
   \mathrm{tr(} \mathcal{M}\mathcal{N})\le 2.
\end{align} 
By using Eq.~\ref{eq:diagonalmatrix}, we have
\begin{align}
   \mathrm{tr(} \mathcal{M}\mathcal{N})   &=2u^2+2[(b^*c-ad^*)(bc^*-a^*d)\\\nonumber
      &+(b^*d - bd^*)(-a^*c + ac^*)]v^2.
\end{align}   

Each term inside the four pairs of parentheses is nothing but the four elements of the matrix $P^{-1}P^*$. By noting that $\det(P^{-1}P^*)=1$ and $u^2+v^2=1$, we have
\begin{align}
      \mathrm{tr(} \mathcal{M}\mathcal{N})&=2u^2+2[1+2(b^*d - bd^*)(-a^*c + ac^*)]v^2\\\nonumber
      &=2u^2+2v^2+4(b^*d - bd^*)(-a^*c + ac^*)v^2\\\nonumber
      &=2+4(b^*d - bd^*)(-a^*c + ac^*)v^2.
\end{align} Hence, $  \mathrm{tr(} \mathcal{M}\mathcal{N}){\le} 2$ implies $(b^*d - bd^*)(-a^*c + ac^*){\le} 0$.
\end{proof}

In our disordered system, we consider the dipolar transfer matrices constructed by $\mathcal{M}=T_+T_-, \mathcal{N}=T_-T_+$ with
\begin{align}
    T_{\pm} =\begin{pmatrix}
E\pm iV  &-1 \\
 1 &0
\end{pmatrix},
\end{align}
leading to
\begin{align}
    \mathcal{M}&=\begin{pmatrix}
 -1 + E^2 + V^2  &-E - i V \\
E - i V &-1
\end{pmatrix}
  ,\\
  \mathcal{N}&=\begin{pmatrix}
 -1 + E^2 + V^2  &-E + i V \\
E + i V &-1
\end{pmatrix},
\end{align}
which satisfies Condition $1$ by construction.

Condition 3 leads to the parametric constraint, $- 4E^2 + E^4 + 2E^2 V^2 + V^4\le 0$, corresponding to the region inside the black line as shown in Fig.~\ref{Fig:PWmapping+Percentagerealspectra+IPRdiagram}. It indeed naturally satisfies Condition 2, which corresponds to the region $E^2+V^2\in [0,4]$. Therefore, for our protocol, we only require Condition 3  $\mathrm{tr}(\mathcal{M}\mathcal{N})\le 2$.

 \let\addcontentsline\oldaddcontentsline
	\cleardoublepage
	\onecolumngrid
 \begin{center}
\textbf{\large{\textit{Supplementary Material} \\ \smallskip
	for ``Non-Hermitian delocalization in 1D via emergent compactness" }}\\
		\hfill \break
		\smallskip
	\end{center}
	
	\renewcommand{\thefigure}{S\arabic{figure}}
	\setcounter{figure}{0}
	\renewcommand{\theequation}{S.\arabic{equation}}
	\setcounter{equation}{0}
	\renewcommand{\thesection}{SM\;\arabic{section}}
	\setcounter{section}{0}
	\tableofcontents

\section{Fraction of delocalized states with real energy and finite-size scaling}
\subsection{Fraction of delocalized states}~\label{SM:percentage}
\begin{figure}[h]
\centering
\includegraphics[width=0.35\linewidth]{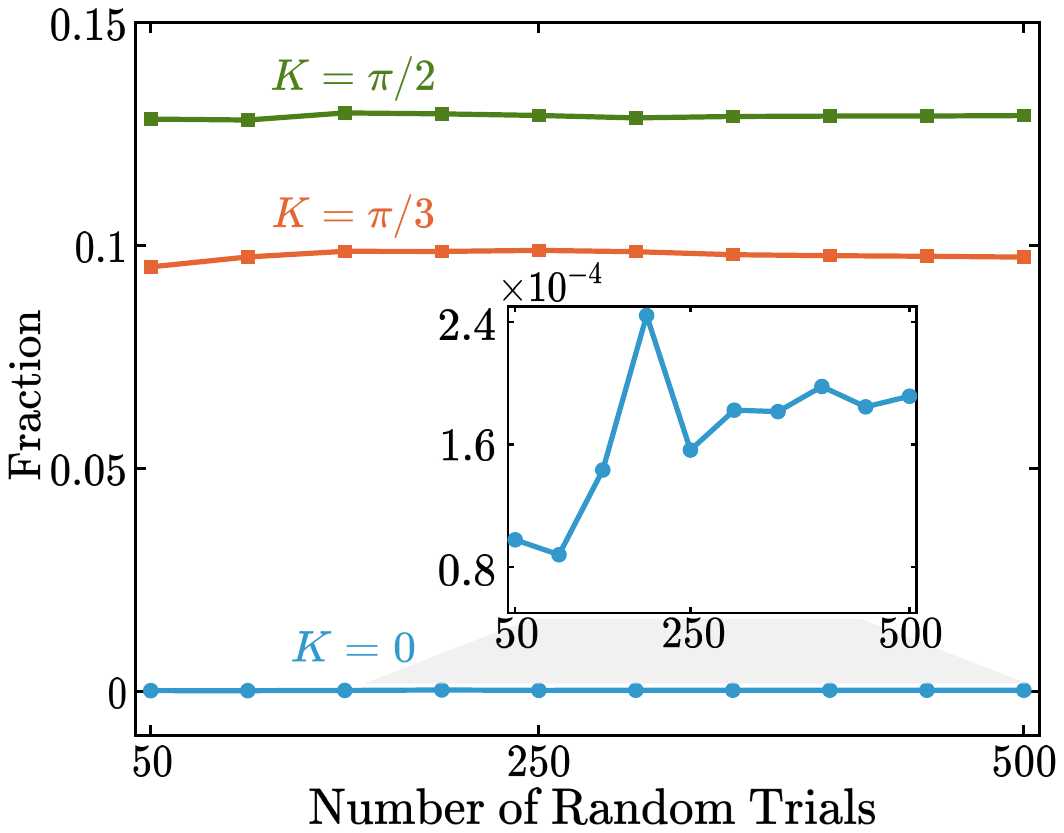}
\caption{Fraction of the delocalized eigenstates with real energy. The fraction quickly converges for an increasing number of random realizations. For the periodic boundary condition, $K=0$, this fraction is very small but non-zero. Here we use $V=0.3$ and system size $1024$, the same as Fig. 1(c) in the main text.}
\label{Fig:percentage}
\end{figure}

The fraction of delocalized eigenstates depends on the specific spatial random realization. 
In Fig.~1 (c) in the main text, the fraction is obtained by averaging over a large number of random realizations. This fraction quickly converges when we increase the size of the random ensemble, as shown in Fig.~\ref{Fig:percentage}. In particular, for the periodic boundary condition $K=0$ (blue), the fraction is around $10^{-4}$, a very small but non-vanishing value.

\subsection{Finite-size scaling}~\label{SM:finitesize}
\begin{figure}[h]
\centering
\includegraphics[width=0.35\linewidth]{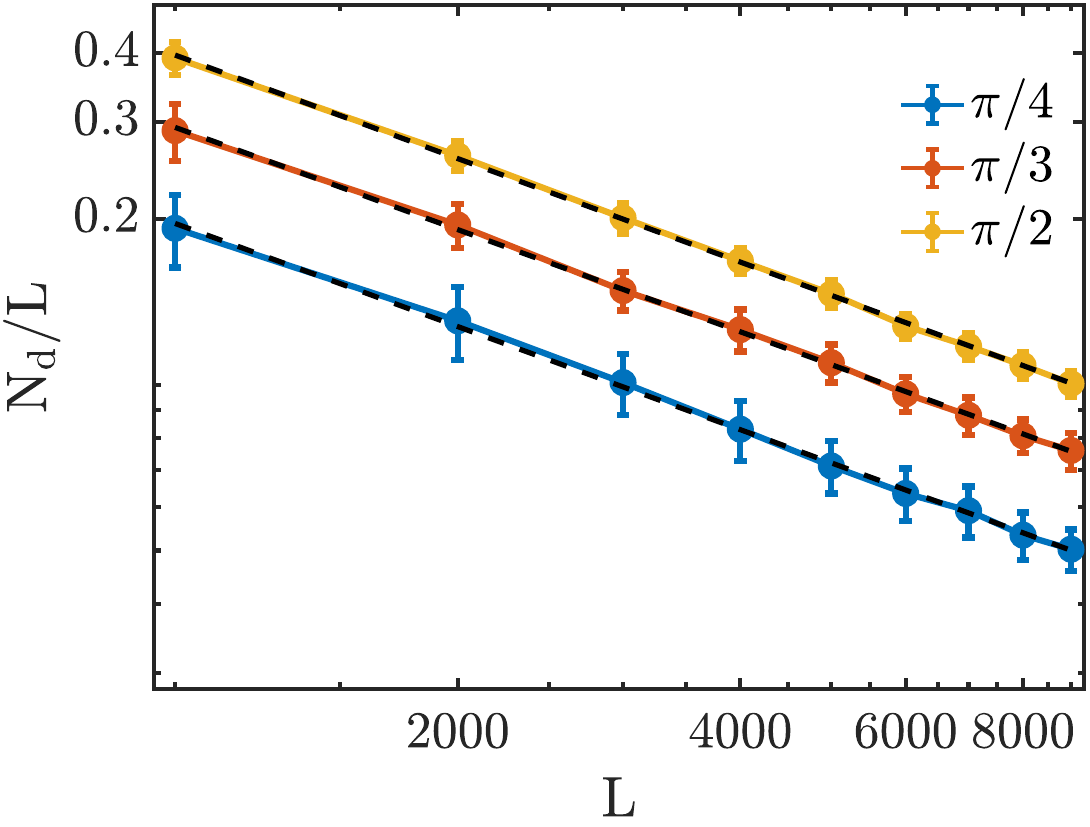}
\caption{Finite-size scaling of the fraction of delocalized eigenstates, averaged over $100$ random disorder realizations for different $K$. Error bars indicate statistical variations. The plot uses a log-log scale and the dashed lines of slope $-0.6$ are plotted as a guide to the eye. We set $V=0.1$ for numerical simulation. }
\label{Fig:Real_Delocal_Fraction}
\end{figure}

We further investigate the finite-size scaling of the total number, $N_{\mathrm{d}}$, and the fraction of the delocalized eigenstates, $N_{\mathrm{d}}/L$. For general boundary condition \(K\), our numerical simulation shows that $N_{\mathrm{d}}$ increases with system size following a power law scaling \(L^{\alpha}\), hence $N_{\mathrm{d}}$ diverges in the thermodynamic limit. In contrast, for $K=0$ and $\pi$, $N_{\mathrm{d}}$ remains small and does not increase for larger $L$.
As shown in Fig.~2(b) in the main text, for \(V=0.1\), \(\alpha=0.4\). The fraction of delocalized eigenstates reads \(N_{\mathrm{d}}/L\propto L^{\alpha-1}\). Since $\alpha<1$, this fraction decays to zero in the thermodynamic limit. As shown in Fig.~\ref{Fig:Real_Delocal_Fraction}, the fraction decays as $L^{-0.6}$. 

\begin{figure}[h]
\centering
\includegraphics[width=0.35\linewidth]{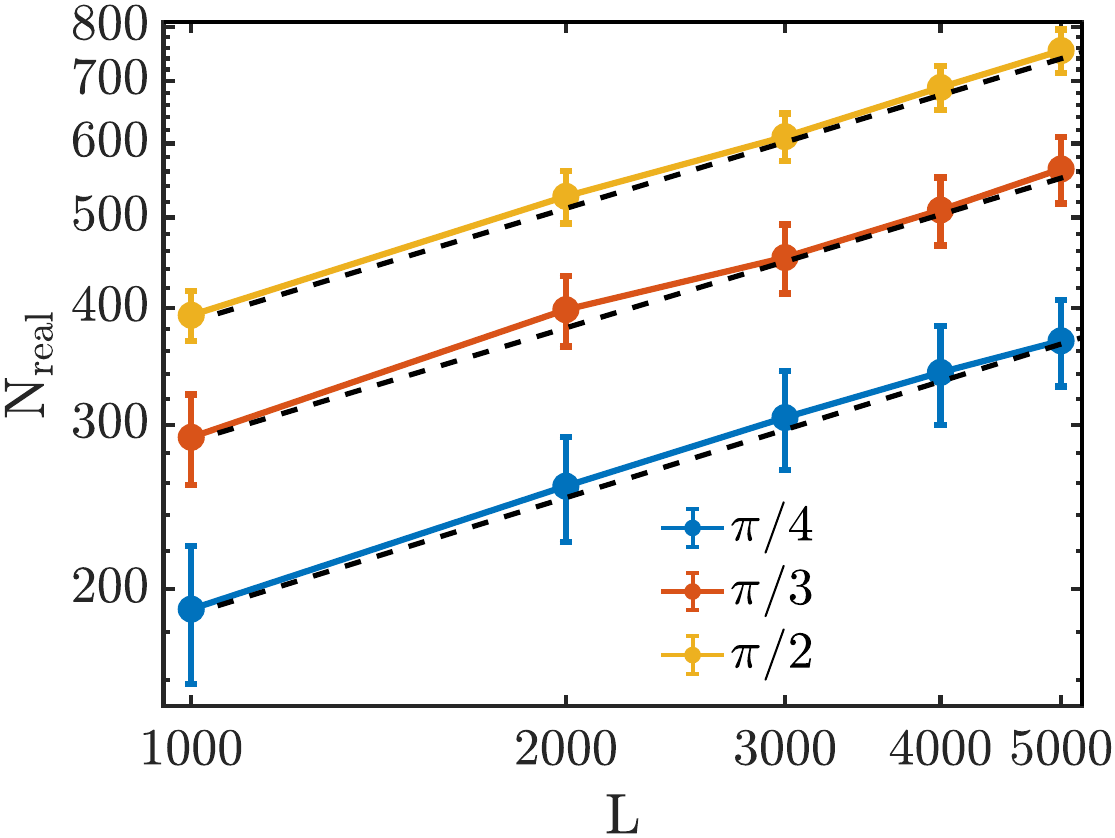}
\caption{Finite-size scaling of the number of real energies, averaged over $100$ trials for different $K$. Error bars indicate statistical variations. The plot uses a log-log scale with dashed lines of slope $0.4$. The scaling behavior indicates that \(N_{\mathrm{real}}\) follows the same power-law growth with system size as $N_{\mathrm{d}}$. Here, we set \( V = 0.1 \). }
\label{Fig:realcountsscaling}
\end{figure}

Moreover, Fig.~\ref{Fig:realcountsscaling} reveals that the total number of real eigenvalues, including both delocalized and localized eigenstates, \(N_{\mathrm{real}}\), also follows a power-law scaling with an exponent of \(0.4\) for \(V=0.1\). This observation implies that the ratio between $N_{\mathrm{real}}$ and $N_{\mathrm{d}}$ remains constant.

\begin{figure}[h]
\centering
\includegraphics[width=0.6\linewidth]{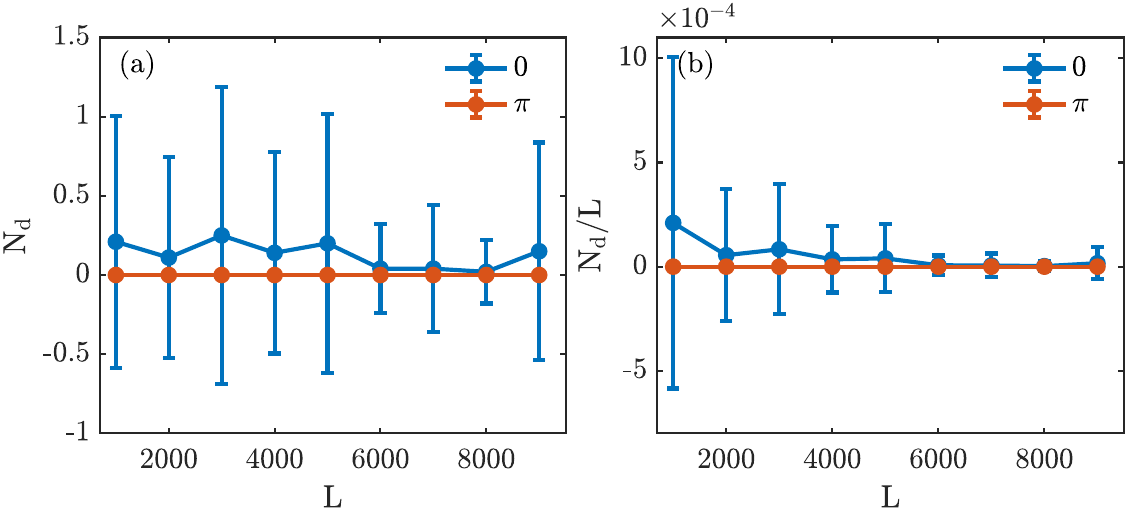}
\caption{(a) Finite-size scaling of $N_\mathrm{d}$ for $K=0,\pi$, averaged over $100$ random disorder realizations. $N_\mathrm{d}$ remains small and largely independent of system sizes. This behavior is notably different from the power-law scaling for other $K$ values. We use \( V = 0.1 \) for numerical simulation.  (b)  $N_{\mathrm{d}}/L$ quickly vanishes for larger system sizes. Error bars correspond to the statistical variations, which can be even larger than the averaged values.}
\label{Fig:finitesizebdyK}
\end{figure}

For \(K=0,\pi\), Fig.~\ref{Fig:finitesizebdyK} shows that $N_{\mathrm{d}}$ remains small and largely independent of the system size. This behavior is notably different from the power-law scaling for generic $K$ values. The fraction of delocalized eigenstates $N_{\mathrm{d}}/L$ also quickly decays. This small number is consistent with the small fraction observed in Fig.~\ref{Fig:percentage}.

\begin{figure}[h]
\centering
\includegraphics[width=0.6\linewidth]{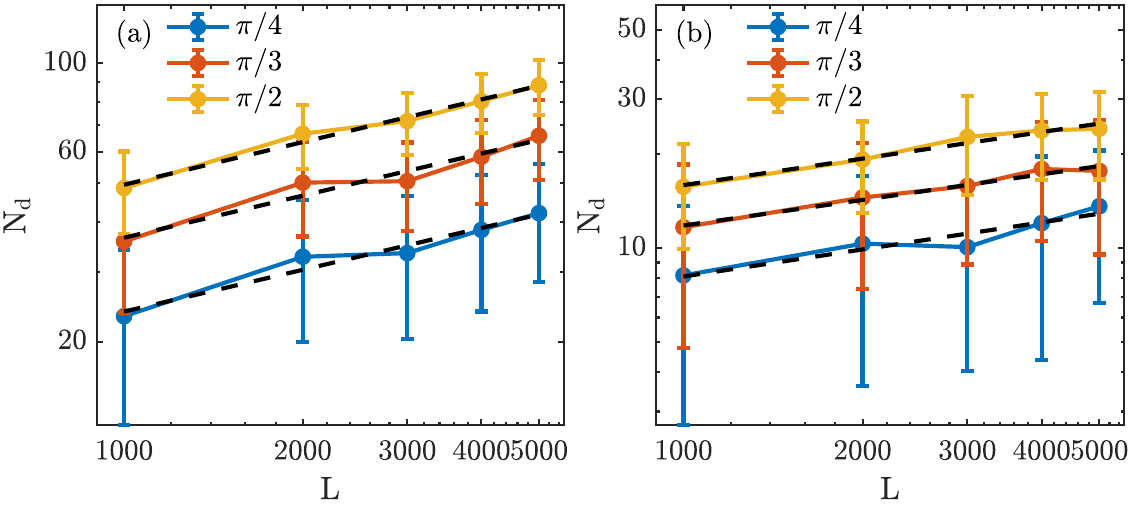}
\caption{Finite-size scaling of $N_{\mathrm{d}}$ averaged over $100$ trials. Error bars represent statistical fluctuations, and different colors correspond to distinct \(K\) values. A power-law dependence is observed. In panel (a), \( V=0.6 \) and the scaling exponent  $\alpha$ is about 0.35,  while in panel (b) with \( V=0.9 \), the exponent is around 0.28.
}
\label{Fig:diffVscalbdy}
\end{figure}

Note, the exponent depends on the disorder strength $V$, typically ranging from \(0.2\) to \(0.4\) based on our numerical observation. In Fig.~\ref{Fig:diffVscalbdy}, we plot $N_{\mathrm{d}}$ obtained for $V=0.6$ and $0.9$ in panels (a) and (b) respectively. The corresponding power-law scaling exponent $\alpha=0.35$ and 0.28. 

\section{$D_2$ symmetry of the spectrum }\label{Sec:symmetry}

In this section, we show that the spectrum has a $D_2$ symmetry, namely, the mirror symmetry w.r.t. both real and imaginary axes.

\subsection{Chiral symmetry}
We first focus on the chiral symmetry, which ensures the coexistence of the energies $E$ and $-E^*$ in the spectrum. 
Consider the single-particle Hamiltonian matrix $H$ of Eq.~\ref{eq:Hamiltonian}. We now show that this Hamiltonian satisfies the following chiral symmetry
\begin{align}
    PH^{\dagger}P^{-1}=-H,
\end{align}
where 
\begin{align}
    P=\begin{pmatrix}
 1 &  &  &  &  &  &  & \\
  & -1 &  &  &  &  &  & \\
  &  & 1 &  &  &  &  & \\
  &  &  & -1  &  &  &  & \\
  &  &  &  & \ddots &  &  & \\
  &  &  &  &  & -1 &  & \\
  &  &  &  &  &  &1  & \\
  &  &  &  &  &  &  &-1
\end{pmatrix}.
\end{align}
\begin{proof}
 \( H \) can be decomposed into a diagonal term and a non-diagonal term:
\[
H = H_{\text{diag}} + H_{\text{nondiag}}.
\]
Under Hermitian conjugation, the Hamiltonian transforms as:
\[
H^\dagger = -H_{\text{diag}} + H_{\text{nondiag}}.
\]
Applying the transformation \( P \), the diagonal term \( H_{\text{diag}} \) satisfies:
\[
P H_{\text{diag}}^\dagger P^{-1} = -P H_{\text{diag}} P^{-1} = -H_{\text{diag}}.
\]
The nearest-neighbor hopping term \( H_{\text{nondiag}} \) anti-commutes with \( P \), leading to:
\[
P H_{\text{nondiag}} P^{-1} = -H_{\text{nondiag}}.
\]
Consequently, the full Hamiltonian under the transformation \( P \) satisfies:
\[
P H^\dagger P^{-1} = -H.
\]
\end{proof}
This symmetry ensures that the energy eigenvalues of $H$ appear in pairs $\{E,-E^*\}$. To see this, suppose that $|\psi\rangle$ is an eigenstate of $H$
\begin{align}
    H|\psi\rangle=E|\psi\rangle.
\end{align}
Applying the symmetry operator $P$ on each side, we have
\begin{align}
&P    HP^{-1}P|\psi\rangle=PE|\psi\rangle\\
\Rightarrow&-H^{\dagger}P|\psi\rangle=EP|\psi\rangle\\
\Rightarrow&H^{\dagger}P|\psi\rangle=-EP|\psi\rangle
\end{align}
Take the Hermitian conjugate for each side, denote $(|\psi\rangle)^{\dagger}=\langle\psi'|$,  we obtain
\begin{align}
   \langle\psi'|P^{\dagger}H=-\langle\psi'|P^{\dagger}E^*.
\end{align}
Therefore, $-E^*$ is also an eigenvalue for the Hamiltonian with the left eigenstate being $\langle\psi'|P^{\dagger}$. 

In fact, these results hold for any random disordered potential that is purely imaginary, not limited to the specific case of our dipolar potential.
\subsection{Symmetric energy pairs \texorpdfstring{$\pm E$}{±E} in the spectrum}\label{SM:Esquare}
We now prove that the solutions of the eigenvalue problem of the transfer matrix (Eq.~\ref{eq:bulk-boundary-equation2}) with twisted boundary condition comes into pairs with eigenvalues $\pm E$. This is achieved by showing that the trace of any random multiplication of $\mathcal{M}$ and $\mathcal{N}$ is always a polynomial of $E^2$. 
\begin{proof}
The explicit form of matrices $\mathcal{M}$ and $\mathcal{N}$ are
\begin{align}
    \mathcal{M}=\begin{pmatrix}
 -1 + E^2 + V^2  &-E - i V \\
E - i V &-1
\end{pmatrix}
  , \mathcal{N}=\begin{pmatrix}
 -1 + E^2 + V^2  &-E + i V \\
E + i V &-1
\end{pmatrix}.
\end{align}
We expand them into Pauli matrices 
\begin{align}
\frac{-2+E^2+V^2}{2}\sigma_0+ \frac{E^2+V^2}{2}\sigma_3\pm iV\sigma_1-iE\sigma_2\label{Eq:stucture}\equiv G_0\sigma_0+G_3\sigma_3\pm G_1\sigma_1+G_2\sigma_2,
\end{align}
where $-$ and $+$ correspond to $\mathcal{M}$ and $\mathcal{N}$, respectively.  Here $G_{i=0,1,2,3}$ denote the coefficient of the four matrices. Crucially, we note that $G_0$ and $G_3$ are polynomial functions of $E^2$ and even powers of $G_{i=1,2}$ are monomials of $E^2$. For example, $G_1^2$ is $V^2$, which is a monomial of zero power of $E^2$. This observation turns out to be important for our argument, as further demonstrated below.

Now we consider a random product of $\mathcal{M}$ and $\mathcal{N}$, e.g.,
\begin{eqnarray}
\mathcal{M}_1\mathcal{M}_2\mathcal{N}_3\mathcal{M}_4\dots\mathcal{N}_{\alpha}\dots\mathcal{M}_k,
\end{eqnarray}
where we use the subindex $\alpha = 1,2,3,4\dots$ to track the position of each matrix. In terms of Pauli matrices, it becomes
\begin{eqnarray}
\label{eq:products}
    \sum_{i_1=0}^3C^{1}_{i_1} \sigma_{i_1}\sum_{i_2=0}^3C^2_{i_2} \sigma_{i_2}\dots \sum_{i_{\alpha}=0}^3C^{\alpha}_{i_{\alpha}} \sigma_{i_{\alpha}}\dots\sum_{i_k=0}^3C^{k}_{i_k} \sigma_{i_k}.
\end{eqnarray}
According to Eq.~\ref{Eq:stucture}, randomness only occurs in $C_1^{\alpha}$ which can take values of $\pm iV$, and $C_{0,2,3}^{\alpha}$ are fixed for all $\alpha$.  The expanded result of the polynomial multiplication is a linear combination of $k$-product of Pauli matrices, e.g.,
\begin{align}
   \sum_{\alpha=1}^k \sum_{i_{\alpha}=0}^3C^{1}_{i_1}\sigma_{i_1}C^2_{i_2}\sigma_{i_2}\dots C^{\alpha}_{i_{\alpha}}\sigma_{i_{\alpha}} \dots C^{k}_{i_k}\sigma_{i_k}.\label{Eq:polyexpand}
\end{align}
We are going to prove that the summation of these $4^k$ terms, each being the $k$-product of Pauli strings, always has a trace that is a polynomial function of $E^2$. 

Utilizing the (anti-)commutation relations among Pauli matrices, each element in Eq.~\ref{Eq:polyexpand} can be simplified as follows:
\begin{equation}
  C^{1}_{i_1}\sigma_{i_1}C^2_{i_2}\sigma_{i_2}\dots C^{\alpha}_{i_{\alpha}}\sigma_{i_{\alpha}} \dots C^{k}_{i_k}\sigma_{i_k}=c_0  (G_0\sigma_0)^t(G_1\sigma_1)^p(G_2\sigma_2)^q(G_3\sigma_3)^l
    =c_0  G_0^t(G_1\sigma_1)^p(G_2\sigma_2)^q(G_3\sigma_3)^l\label{Eq:paulistringsimplify},
\end{equation}
where $t,p,q,l$ refer to the total number of different Pauli matrices corresponding to a given string $\{i_1, i_2,\dots, i_k\}$, satisfying $p+q+l=k-t$. The overall prefactor $c_0$ is either $1$ or $-1$,  originating from the anti-commutation of two Pauli matrices as well as the randomness of the appearance of  $\sigma_1$ in $\mathcal{M}$ and $\mathcal{N}$.

A crucial observation is that any string $\{i_1, i_2, \dots, i_k\}$ containing at least one $\sigma_2$ and one $\sigma_3$ must have its unique counterpart in the expansion given by Eq.~\ref{Eq:polyexpand}. This counterpart is identified by locating the first pair of $\sigma_2$ and $\sigma_3$ from left to right in the sequence and swapping these two elements, regardless of whether they are adjacent. These two sequences share identical coefficients, but their permutation parities are opposite, resulting in mutual cancellation.

To illustrate, we can count the occurrences of $\sigma_2$ and $\sigma_3$ from left to right across the string
and track their cumulative numbers. We define a number $n$ as the position in the string where, for the first time, we encounter one $\sigma_2$ and $\sigma_3$, such that the elements before $n$ include either zero $\sigma_2$ or zero $\sigma_3$.

If the $n$th term is $\sigma_3$, then we swap it with the nearest $\sigma_2$ in the preceding $n-1$ terms; if the $n$th term is $\sigma_2$, then we swap it with the nearest $\sigma_3$ in the former $n-1$ terms. After the swap, the resulting string will be a unique counterpart of the original. 

This term naturally exists in the expansion (Eq.~\ref{Eq:polyexpand})  because of the distributive property in the polynomial multiplication.

For example, consider the string $\{ i_1=1, i_2=2, i_3=1,i_4=3, i_5=2 \}$ with $k=5$. We proceed with the following steps:
\begin{itemize}
    \item Counting each term, we notice:\begin{align*}
        &C^{1}_{i_1=1}\sigma_1 C^{2}_{i_2=2}\sigma_2 \quad &\text{(1 $\sigma_2$, 0 $\sigma_3$)} \\
        &C^{1}_{i_1=1}\sigma_1C^{2}_{i_2=2} \sigma_2C^{3}_{i_3=1}\sigma_1 \quad &\text{(1 $\sigma_2$, 0 $\sigma_3$)} \\
        &C^{1}_{i_1=1}\sigma_1C^{2}_{i_2=2} \sigma_2C^{3}_{i_3=1}\sigma_1 C^{4}_{i_4=3}\sigma_3 \quad &\text{(1 $\sigma_2$, 1 $\sigma_3$)}
    \end{align*}
    \item In the fourth term, both a $\sigma_2$ and a $\sigma_3$ are present, meeting the criteria set forth above. Thus, we swap the $\sigma_3$ at the fourth position with the nearest $\sigma_2$, which is at the second position.
    \item The resultant string after the swap is $\{ i_1=1, i_2=3, i_3=1,i_4=2, i_5=2 \}$ and its explicit form is $C^{1}_{i_1=1}\sigma_1C^{2}_{i_2=2} \sigma_3C^{3}_{i_3=1}\sigma_1 C^{4}_{i_4=3}\sigma_2C^{5}_{i_5=2}\sigma_2$, which is contained in the expansion of the $k=5$-terms polynomial multiplication in Eq.~\ref{Eq:polyexpand}. Also, it has the same coefficient, $C^{1}_{i_1=1}C^{2}_{i_2=2} C^{3}_{i_3=1} C^{4}_{i_4=3}C^{5}_{i_5=2}$, as the original string.
\end{itemize}

Now we demonstrate that swapping the nearest $\sigma_2$ and $\sigma_3$ in the sequence results in opposite permutation parities. The closest pair of $\sigma_2$ and $\sigma_3$ appears in two ways, either with an even or odd number of $\sigma_1$ matrices in between. In the even case, this simplifies to the swapping of the first adjacent pair of $\sigma_2$ and $\sigma_3$, which naturally yields opposite permutation parities. For the other case, one only needs to consider the permutation parities of $\sigma_2\sigma_1\sigma_3$ and $\sigma_3\sigma_1\sigma_2$, which are also opposite.

Therefore, any string $\{i_1, i_2,\dots, i_k\}$ that contains at least one \(\sigma_2 \) and one \( \sigma_3 \) does not contribute to the trace of Eq.~\ref{eq:products}. We only need to study Pauli strings without $\sigma_2$ and $\sigma_3$ appearing at the same time, i.e. either $q=0$ or $l=0$. 

\textbf{Case 1:} $l=0$

In this case, if both $p$ and $q$ are even, the trace is a polynomial of $E^2$. Conversely, if either $p$ or $q$ is odd, the trace vanishes. Specifically:
\begin{itemize}
    \item If only one of $p$ or $q$ is odd, then the trace of Eq.~\ref{Eq:paulistringsimplify} becomes proportional to either $\mathrm{tr}(\sigma_{1})$ or $\mathrm{tr}(\sigma_{2})$, which vanishes.
    \item If both $p$ and $q$ are odd, the trace is proportional to $\mathrm{tr}(\sigma_{3})$, which is also zero.
\end{itemize}

\textbf{Case 2:} $q=0$

The analysis mirrors that of Case 1, where terms with even $p$ and $l$ survive, and their traces are polynomials of $E^2$.

In summary, only terms with even powers of $\sigma_1$ and $\sigma_2(\sigma_3)$ contribute to the trace of Eq.~\ref{Eq:polyexpand}, thus the trace is a polynomial of $E^2$.
\end{proof}

In conclusion, we prove that the trace of any random multiplication of $\mathcal{M}$ and $\mathcal{N}$ is a polynomial of $E^2$. In other words, $E$ and $-E$ must coexist in the spectrum. This symmetry and the chiral symmetry result in the $D_2$ symmetry in the spectrum.

\end{document}